\documentclass[conference]{IEEEtran}

\usepackage{times}
\usepackage{soul}
\usepackage{url}
\usepackage[hidelinks]{hyperref}
\usepackage[utf8]{inputenc}
\usepackage[small]{caption}
\usepackage{graphicx}
\usepackage{amsmath}
\usepackage{amsthm}
\usepackage{booktabs}
\usepackage{algorithm}
\usepackage{algorithmic}
\usepackage{multirow}
\newtheorem{example}{Example}
\newtheorem{theorem}{Theorem}
\newtheorem{definition}{Definition}
\newtheorem{remark}{Remark}
\newtheorem{lemma}{Lemma}

\usepackage{stix,amsbsy} 
\usepackage{tikz}
\usepackage[geometry]{ifsym}
\usetikzlibrary{shapes,shadows,arrows,decorations.pathreplacing, positioning}

\definecolor{niceblue}{rgb}{0.0, 0.4, 0.6}

\DeclareMathOperator\supp{supp}

\newcommand{\s}{\bar{s}}
\newcommand{\up}{\texttt{up}}
\newcommand{\lt}{\texttt{left}}
\newcommand{\dn}{\texttt{down}}
\newcommand{\rt}{\texttt{right}}

\usepackage{xspace}
\newcommand{\A}{{\cal A}}

\newcommand{\M}{{\cal M}}

\newcommand{\T}{{\cal T}}

\newcommand{\Wiz}{\text{\sc Wiz}\xspace}
\newcommand{\MB}{\text{\sc MB}\xspace}
\newcommand{\true}{\texttt{true}\xspace}

\renewcommand{\path}{\text{path}\xspace}
\newcommand{\pred}{\text{pred}\xspace}

\newcommand{\PO}{Player~$1$\xspace}
\newcommand{\PT}{Player~$2$\xspace}
\newcommand{\PLi}{Player~$i$\xspace}

\newcommand{\G}{{\cal G}}

\newcommand{\Q}{\mathbb{Q}}
\newcommand{\Nat}{\mbox{I$\!$N}}
\newcommand{\Real}{\mbox{I$\!$R}}
\newcommand{\stam}[1]{}
\newcommand{\zug}[1]{\left( #1  \right)}
\newcommand{\set}[1]{\{ #1 \}}

\usepackage[draft]{todonotes} 

\title{Formal Methods with a Touch of Magic}
\author{
\IEEEauthorblockN{Parand Alizadeh Alamdari}
\IEEEauthorblockA{Sharif University of Technology}
\and
\IEEEauthorblockN{Guy Avni\\  Thomas A. Henzinger\\  Anna Lukina}
\IEEEauthorblockA{IST Austria}
}

\begin{document}

\maketitle

\begin{abstract}
Machine learning and formal methods have complimentary benefits and drawbacks.   In this work, we address the controller-design problem with a combination of techniques from both fields.  The use of black-box neural networks in deep reinforcement learning (deep RL) poses a challenge for such a combination.  Instead of reasoning formally about the output of deep RL, which we call the {\em wizard}, we extract from it a decision-tree based model, which we refer to as the {\em magic book}. Using the extracted model as an intermediary, we are able to handle problems that are infeasible for either deep RL or formal methods by themselves. First, we suggest, for the first time, a synthesis procedure that is based on a magic book. We synthesize a stand-alone correct-by-design controller that enjoys the favorable performance of RL. Second, we incorporate a magic book in a bounded model checking (BMC) procedure. BMC allows us to find numerous traces of the plant under the control of the wizard, which a user can use to increase the trustworthiness of the wizard and direct further training.
\end{abstract}

\section{Introduction}
Machine-learning techniques and, in particular, the use of neural networks (NNs), are exploding in popularity and becoming a vital part of the development of many technologies. There is a challenge, however, in deploying systems that use trained components, which are inherently black-box.
For a system to be used by a human, it must be trustworthy: provably correct, or predictable, at the least. Current trained systems lack either of these properties. %Combining machine learning with formal methods is a promising approach to develop trustworthy systems. 

In this work, we focus on the controller-design problem. Abstractly speaking, a controller is a device that interacts with a plant. At each time step, the plant outputs its state and the controller feeds back an action. Combining techniques from both formal methods and machine learning is especially appealing in the controller-design problem since it is critical that the designed controller is both correct and that it optimizes plant performance.
%The goal is to construct a provably correct controller that optimizes plant performance.
%The problem of controller design has been addressed separately in both communities. 
%Solutions that combine techniques from both fields are particularly appealing, as both performance and safety are critical.

{\em Reinforcement learning} (RL) is the main machine-learning tool for designing controllers. The RL approach is based on ``trial and error'': the agent randomly explores its environment, receives rewards and learns from experience how to maximize them. RL has made a quantum leap in terms of scalability since the recent introduction of NNs into the approach, termed {\em deep RL} \cite{MK+13}. We call the output of deep RL the {\em wizard}: it optimizes plant performance but, since it is a NN, it does not reveal its decision procedure. More importantly, there are no guarantees on the wizard and it can behave unexpectedly and even incorrectly. 

Reasoning about systems that use NNs poses a challenge for formal methods. First, in terms of scalability (NNs tend to be large), and second, the operations that NNs depend on are challenging for formal methods tools, namely NNs use numerical rather than Boolean operations and {\em ReLu} neurons use the \texttt{max} operator, which SMT tools struggle with. 

We propose a novel approach based on extracting a decision-tree-based model from the wizard, which approximates its operation and is intended to reveal its decision-making process. Hence, we refer to this model as the {\em magic book}. Our requirements for the magic book are that it is (1) simple enough for formal methods to use, and (2) a good approximation of the NN. 

Extracting decision-tree-based models that approximate a complicated function is an established practice \cite{ernst2005tree}. The assumption that allows this extraction to work is that a NN contains substantial redundancy. During training, the NN ``learns'' heuristics that it uses to optimize plant performance. The heuristics can be compactly captured in a small model, e.g., in a decision-tree. This assumption has led, for example, to attempts of distilling knowledge from a trained NN to a second NN during its training~\cite{HVD15,frosst2017distilling}, and of minimizing NNs (e.g., \cite{shriver2019refactoring}). The extraction of a simple model is especially common in {\em explainable AI} (XAI) \cite{Adadi2018PeekingIT}, where the goal is to explain the operation of a learned system to a human user. 

\medskip
We use the tree-based magic book to solve problems that are infeasible both for deep RL and for formal methods alone. Specifically, we illustrate the magic book's benefit in two approaches for designing controllers as we elaborate below.
%: first, we incorporate for the first time, a magic book into a {\em reactive synthesis} \cite{PR89} procedure to produce stand-alone correct-by-design controllers, where the magic book is first used to add performance considerations and then to relax the adversarial assumption on the environment in a multi-agent setting. Second, we devise a {\em bounded model checking} (BMC) procedure \cite{} based on a magic book and use to efficiently generate a large number of corner-case traces, which we use in verification and explainability of the wizard to increase its trustworthiness.

\medskip
{\em Reactive synthesis} \cite{PR89} is a formal approach to design controllers. The input is a qualitative specification and the output is a correct-by-design controller. The fact that the controller is provably correct, is the strength of synthesis. A first weakness of traditional synthesis is that it is purely qualitative and specifications cannot naturally express quantitative performance. There is a recent surge of quantitative approaches to synthesis (e.g., \cite{BCHJ09,BB+13,AK+17}). However, these approaches suffer from other weaknesses of synthesis: deep RL vastly outperforms synthesis in terms of scalability. Also, in the average-case, RL-based controllers beat synthesized controllers since the goal in synthesis is to maximize worst-case performance. 

Synthesis is often reduced to solving a two-player graph game; \PO represents the controller and \PT represents the plant. In each step, \PT reveals the current state $\s$ of the plant and \PO responds by choosing an action. In our construction, when \PT chooses $\s$, we extract from the magic book the action $a$ that is taken at $\s$. \PO's action then depends on $a$ as we elaborate below. The construction of the game arena thus depends on the magic book, and using the wizard instead is infeasible. 

We present a novel approach for introducing performance considerations into reactive synthesis. We synthesize a controller that satisfies a given qualitative specification while following the magic book as closely as possible. We formalize the later as a quantitative objective: whenever \PO agrees with the choice of action suggested by \PT, he receives a reward, and the goal is to maximize rewards. Since the magic book is a proxy for the RL-generated wizard, we obtain the best of both worlds: a provably correct controller that enjoys the high average-case performance of RL. In our experiments, we synthesize a controller for a taxi that travels on a grid for the specification ``visit a gas station every $t$ steps'' while following advice from a wizard that is trained to collect as many passengers as possible in a given time frame. 

In a second application, we use a magic book to relax the adversarial assumption on the environment in a multi-agent setting. We are thus able to synthesize controllers for specifications that are otherwise {\em unrealizable}, i.e., for which traditional synthesis does not return any controller. Our goal is to synthesize a controller for an agent that interacts with an environment that consists of other agents. Instead of modeling the other agents as adversarial, we assume that they operate according to a magic book. This restricts their possible actions and regains realizability. For example, suppose a taxi that is out of our control, shares a network of roads with a bus, under our control. Our goal is to synthesize a controller that guarantees that the bus travels between two stations without crashing into a taxi. While an adversarial taxi can block the bus, by assuming that the taxi operates according to a magic book, we limit \PT's action in the game and find a winning \PO strategy that corresponds to a correct controller.

\medskip
{\em Bounded model checking}~\cite{BC+03} (BMC) is an established technique to find bounded traces of a system that satisfy a given specification. In a second approach to the controller-design problem, we use BMC as an XAI tool to increase the trustworthiness of a wizard before outputting it as the controller of the plant. 
We rely on BMC to find (many) traces of the plant under the control of the wizard that are tedious to find manually. 
%A user can use BMC to search for (many) corner-case traces of the plant under the control of the wizard to obtain insights on the operation of the wizard. Then, the user can utilize the insights  in a secondary training of the wizard. 

We solve BMC by constructing an SMT program that intuitively simulates the operation of the plant under the control of the magic book rather than under the control of the wizard. The traces we find witness the magic book. A disadvantage of the approach is that it is not sound (see Remark~\ref{rem:sound}). The advantage is that the reduction from BMC to SMT is simple and leads to a significant performance gain: in our experiments, we use the standard SMT solver Z3~\cite{MouraB08} to extract thousands of witnesses within minutes, whereas Z3 is incapable of solving extremely modest wizard-based BMC instances. Before outputting a trace, we perform a secondary test to check that it witnesses the wizard as well.  In our experiments, we find that many traces are indeed shared between the two. Thus, our procedure efficiently finds numerous traces of the plant under the control of the wizard.

A first application of BMC is in verification; namely, we find counterexamples for a given specification. For example, when controlling a taxi, a violation of a liveness property is an infinite loop in which no passenger is collected. We find it more appealing to use BMC as an XAI tool. For example, BMC allows us to find ``suspicious'' traces that are not necessarily incorrect; e.g., when controlling a taxi, a passenger that is not closest is collected first. Individual traces can serve as explanations. Alternatively, we use BMC's ability to find many traces and gather a large dataset. We extract a small human-interpretable model from the dataset that attempts to explain the wizard's decision-making procedure. For example, the model serves as an answer to the question: when does the wizard prefer collecting a passenger that is not closest?

\subsection{Related work}
We compare our synthesis approach to {\em shielding} \cite{KAB+17,AB+19}, which adds guarantees to a learned controller at runtime by monitoring the wizard and correcting its actions. Unlike shielding, the magic book allows us to open up the black-box wizard, which, for example, enables our controller to cross an obstacle that was not present in training, a task that is inherently impossible for a shield-based controller. A second key difference is that we produce stand-alone controllers whereas a shield-based approach needs to execute the NN wizard in each step. Our method is thus preferable in settings where running a NN is costly, e.g., embedded systems or real time systems.

To the best of our knowledge, synthesis in combination with a magic book was never studied. Previously, finding counterexamples for tree-based controllers that are extracted from NN controllers was studied in  \cite{BPS18} and \cite{TN19}. The ultimate goal in those works is to output a correct tree-based controller. A first weakness of this approach is that, since both wizard and magic book are trained, they exhibit many correctness violations. We believe that repairing them manually while maintaining high performance is a challenging task. Our synthesis procedure assists in automating this procedure. Second, in some cases, a designer would prefer to use a NN controller rather than a tree-based one since NNs tend to generalize better than tree-based models. Hence, we promote the use of BMC for XAI to increase the trustworthiness of the wizard. Finally, the case studies the authors demonstrate are different from ours, thus they strengthen the claim that a tree-based classifier extraction is not specific to our domain rather it is a general concept. 

A specialized wizard-based BMC tool was recently shown in~\cite{KB+19}, thus unlike our approach, there is no need to check that the output trace is also a witness for the wizard. More importantly, their method is ``sound'': if their method terminates without finding a counterexample for bound $\ell \in \Nat$, then there is indeed no violation of length $\ell$. Beyond the disadvantages listed above, the main disadvantage of their approach is scalability, which is not clear in the paper. As we describe in the experiments section, our experience is that a wizard-based BMC implemented in Z3 does not scale.

Our BMC procedure finds traces that witness a temporal behavior of the plant. This is very different from finding {\em adversarial examples}, which are inputs slightly perturbed so that to lead to a different output. Finding adversarial examples and verifying robustness have attracted considerable attention in NNs (for example, \cite{KB+17,GM+18,HKWW17}) as well as in random-forest classifiers (e.g., \cite{EG+19,DAD19}).

Somewhat similar in spirit to our approach is applying {\em program synthesis} to extract a program from a NN~\cite{VC+18,VM+18,ZX+19}, which, similar to the role of the magic book, is an alternative small model for application of formal methods. The main goal in these papers is to extract a magic book (``program'', in their terminology) from a wizard, verify its correctness and use it as the controller for the plant. A key difference from our synthesis approach is that their wizard is trained to satisfy the specification and the challenge is to devise a good approximation for the wizard. Our wizard, however, is trained without consideration for the specification; e.g., neither the gas station nor the obstacles mentioned above are present in training. The challenge is to incorporate the wizard in synthesis to gain both performance and correctness.

Decision trees were previously used to represent, in a succinct, verifiable and explainable manner, a strategy for a controller (e.g., \cite{ABC+19, AJ+20}). The challenge here is to construct a concise controller from a given policy, similar to the works above. A second difference is that the policy is given explicitly and is obtained from an explicit solution to the MDP or game, hence scalability is an inherent limitation of this approach. 

Finally, examples of other combinations of RL with synthesis include works that run an online version of RL (see \cite{JJ+19} and references therein), an execution of RL restricted to correct traces \cite{KPR18}, and RL with safety specifications \cite{WET15}.

\section{Preliminaries}
\paragraph{Plant and controller}
We formalize the interaction between a controller and a plant. The plant is modelled as a {\em Markov decision process} (MDP) which is $\M = \left(S,\s_0, A, R, p\right)$, where $S$ is a finite set of states, $\s_0 \in S$ is an initial configuration of the state, $A$ is a finite collection of actions, $R:S \rightarrow \Real$ is a reward provided in each state, and $p: S \times A \rightarrow [0,1]^S$ is a probabilistic transition function that, given a state and an action, produces a probability distribution over states.  
\begin{example}
\label{ex:plant}
Our running example throughout the paper is a taxi that travels on an $n\times n$ grid and collects passengers. Whenever a passenger is collected, it re-appears in a random location. A state of the plant contains the locations of the taxi and the passengers, thus it is a tuple $\s = \zug{p_0, p_1,\ldots, p_k}$, where for $0 \leq i \leq k$, the pair $p_i = \zug{x_i, y_i}$ is a position on the grid, $p_0$ is the position of the taxi, and $p_i$ is the position of Passenger~$i$. The set of actions is $A = \set{\up,\rt, \dn, \lt}$. The transitions of $\M$ are largely deterministic: given an action $a \in A$, we obtain the updated state $\s'$ by updating the position of the taxi deterministically, and if the taxi collects a passenger, i.e., $p'_0 = p_i$, for some $1 \leq i \leq k$, then the new position of Passenger~$i$ is chosen uniformly at random. 
\end{example}

The controller is a {\em policy}, which prescribes which action to take given the history of visited states, thus it is a function $\pi: S^* \rightarrow A$. A policy is {\em positional} if the action that it prescribes depends only on the current position, thus it is a function $\pi: S \rightarrow A$. We are interested in finding an optimal and correct policy as we define below.

\paragraph{Qualitative correctness}
We consider a strong notion of qualitative correctness that disregards probabilistic events, often called {\em surely} correctness. A specification is $\Omega \subseteq S^\omega$. We define the {\em support} of $p$ at $\s$ given $a \in A$ as $\supp(\s, a) = \set{\s': p\left (\bar{s}^{\prime}\;\middle\vert\;\bar{s}, a \right ) > 0}$ and, for a policy $\pi$, we define the support of $\pi$ to be $\supp_\pi(\s) = \supp(\s, \pi(\s))$. 
%Examples of specification include safety specifications like ``the taxi never hits a wall'' or liveness specifications like ``if a passenger appears, eventually it gets collected''.
We define the {\em surely language} of $\M$ w.r.t. $\pi$, denoted $L_{\pi}(\M)$. A run $\sigma = \sigma_1, \sigma_2,\ldots \in S^\omega$ is in $L_{\pi}(\M)$ iff we have $\sigma_1 = \s_0$ and for every $i \geq 1$, we have $\sigma_{i+1} \in \supp_\pi(\sigma_i)$, where $a_i = \pi(\sigma_1,\ldots,\sigma_i)$. We say that $\pi$ is surely-correct for plant $\M$ w.r.t. a specification $\Omega \subseteq S^\omega$ iff it allows only correct runs of $\M$, thus $L_\pi(\M) \subseteq \Omega$.

\paragraph{Quantitative performance and deep reinforcement learning}
The goal of reinforcement learning (RL) is to find a policy in an MDP that maximizes the expected reward~\cite{sutton1992reinforcement}. 
In a finite MDP $\M$, the state at a time step $t \in \Nat$ is a random variable, denoted $s_t$. Each time step entails a reward, which is also a random variable, denoted $r_t$. The probability that $s_t$ and $r_t$ get particular values depends solely on the previous state and action. Formally, for an initial state $\s_0 \in S$, we define $\Pr[s_0 = \s_0] = 1$, and for $\bar{s}^{\prime},\bar{s} \in \mathcal{S}$ and $a \in A$, we have $\Pr\left[s_t=\bar{s}^{\prime}, r_t=R(s)\;\middle\vert\; s_{t-1}=\bar{s},a_{t-1}=a\right] = p\left(\bar{s}^{\prime}\;\middle\vert\; \bar{s}, a\right).$
We consider {\em discounted rewards}. Let $\gamma \in (0,1)$ be a discount factor. The {\em expected reward} that a policy $\pi$ ensures starting at state $\s \in S$ is $Rew_\pi(\s) =  \sum_{t = 0}^\infty \gamma^t r_t$, where $r_t$ is defined w.r.t. $\s$ as in the above. The goal is to find the optimal policy $\pi^*$ that attains $\sup_{\pi} Rew_\pi(\s_0)$.

We consider the {\em Q-learning} algorithm for solving MDPs, which relies on a function $Q: S \times A \rightarrow \Real$ such that $Q(\s, a)$ represents the expected value under the assumption that the initial state is $\s$ and the first action to be taken is $a$, thus $Q(\s, a) =  R(\s) + \gamma \cdot \sum_{\s'} p\left(\s' \;\middle\vert\; \s, a\right) \cdot Rew_{\pi^*}(\s')$. Clearly, given the function $Q$, one can obtain an optimal positional policy $\pi^*$, by defining  $\pi^*(\s) = \arg\max_a  Q(\s, a)$, for every state $\s \in S$. In Q-learning, the Q function is estimated and iteratively refined using the Bellman equation. 

Traditional implementations of Q-learning assume that the MDP is represented explicitly. Deep RL~\cite{MK+13} implements the Q-learning algorithm using a symbolic representation of the MDP as a NN. The NN takes as input a state $\s$ and outputs for each $a \in A$, an estimate of $Q(\s, a)$. The technical challenge in deep RL is that it combines training of the NN with estimating the Q function. We call the NN that deep RL outputs the {\em wizard}. Even though deep RL does not provide any guarantees on the wizard, in practice it has shown remarkable success.

\paragraph{Magic books from decision-tree-based classifiers}
Recall that the output of deep RL is a positional function that is represented by a NN $\Wiz: S \rightarrow A$. We are interested in extracting a small function $\MB$ of the same type that approximates $\Wiz$ well. We use {\em decision-tree based classifiers} as our model of choice for $\MB$. Each internal node $v$ of a decision tree is labeled with a predicate $\varphi(v)$ over $S$ and each leaf is labeled with an action in $A$. A plant state $\s$ gives rise to a unique path in a decision tree $\T$, denoted $\path(\T, \s)$, in the expected manner. The first node is the root. Upon visiting an internal node $v$, the next node in $\path(\T, \s)$ depends on the satisfaction value of $\varphi(v)(\s)$. Suppose $\varphi_1, \ldots, \varphi_n$ is the sequence of predicates traversed by a path $\eta = \path(\T, \s)$, we use $\pred(\eta)$ to denote $\varphi_1 \wedge \ldots \wedge \varphi_n$. Thus, for every $\s' \in S$ we have $\eta = \path(\T, \s')$ iff $\s'$ satisfies $\pred(\eta)$. When $\path(\T, \s)$ ends in a leaf labeled $a \in A$, we say that the tree {\em votes} for $a$. A {\em forest} contains several trees. On input $\s \in S$, each tree votes for an action and the action receiving most votes is output. 

To obtain $\MB$ from $\Wiz$, we first execute $\Wiz$ with the plant on a considerable number $T$ of steps, to collect pairs of the form $\left(\bar{s}_t,\Wiz(\bar{s}_t)\right)$, for $t{\in}\{0,\ldots,T\}$, where $\bar{s}_t$ is the system state at time $t$ and $T$ is chosen to maximize model's F1 score. We then employ standard techniques on this dataset to construct either a decision tree, or a forest of decision trees using the state-of-the-art {\em random forest}~\cite{Bre01} or {\em extreme gradient boosting}~\cite{chen2016xgboost} techniques.

\begin{remark}
One might wonder whether it is possible to obtain a decision tree (magic book) directly from RL, thus making the wizard obsolete. While there were attempts at using decision trees as the underlying reward approximation in RL (e.g., \cite{PH01}), the approach has inherent limitations (see details in \url{https://bit.ly/30WBA1i}). Moreover, decision trees are popular data structures that are often preferred to NNs since they are simpler, easier to interpret, and have less parameters to tune. Still, the literature on deepRL overshadows the literature on RL with decision trees. Also, the choice to extract a decision tree from a NN rather than training a decision tree directly was also made in \cite{BPS18,TN19}, and their case studies differ from ours, strengthening the claim that this approach is general. Finally, we note that even if the magic book is obtained directly from RL, it does not solve the challenges we address in our synthesis and BMC procedures.
%While some attempts have been made to use decision trees to succinctly represent a policy \cite{PH01} or to modify Q-learning to obtain smaller size of trees, sometimes at the expense of performance. Extracting a decision-tree controller from a NN was also done in \cite{BPS18,TN19}. The synergy of RL and decision trees is however under-explored. To avoid the risk of sacrificing performance, we argue that the wizard is indeed essential. 
\end{remark}

\section{Synthesis with a Touch of Magic}
Our primary goal in this section is to automatically construct a correct controller and performance is a secondary consideration. We incorporate a magic book into synthesis and illustrate two applications of the constructions that are infeasible without a magic book. 

\subsection{Constructing a game}
\label{sec:game}
Synthesis is often reduced to a two-player graph game (see \cite{BCJ18}). In this section, we describe a construction of a game {\em arena} that is based on a magic book and in the next sections we complete the construction by describing the players' objectives and illustrate applications. In the traditional game, \PT represents the environment and in each turn, he reveals the current location of the plant. \PO, who represents the controller, answers with an action. A strategy for \PO corresponds to a policy (controller) since, given the history of observed plant states,  it prescribes which action to feed in to the plant next. The traditional goal is to find a \PO strategy that guarantees that a given specification $\Omega$ is satisfied no matter how \PT plays. Traditional synthesis is purely qualitative; namely, it returns some correct policy with no consideration to its performance. When no correct controller exists, we say that $\Omega$ is {\em un-realizable}.

Formally, a graph game is played on an arena $\zug{V, \Xi_1, \Xi_2, \delta}$, where $V$ is a set of vertices, for $i \in \set{1,2}$, \PLi's possible actions are $\Xi_i$, and $\delta: V \times \Xi_1 \times \Xi_2 \rightarrow V$ is a deterministic transition function. The game proceeds by placing a token on a vertex in $V$. When the token is placed on $v \in V$, \PT moves first and chooses $\xi_2 \in \Xi_2$. Then, \PO chooses $\xi_1 \in \Xi_1$ and the token proceeds to $\delta(v, \xi_1, \xi_2)$. In games, rather than using the term ``policy'', we use the term {\em strategy}. Two strategies $f$ and $g$ for the two players and an initial vertex induce a unique infinite play, which we denote by $\textit{play}(f, g)$, where for ease of notation we omit the initial vertex. 

\begin{figure}[ht]
\centering
\includegraphics[height=2.5cm]{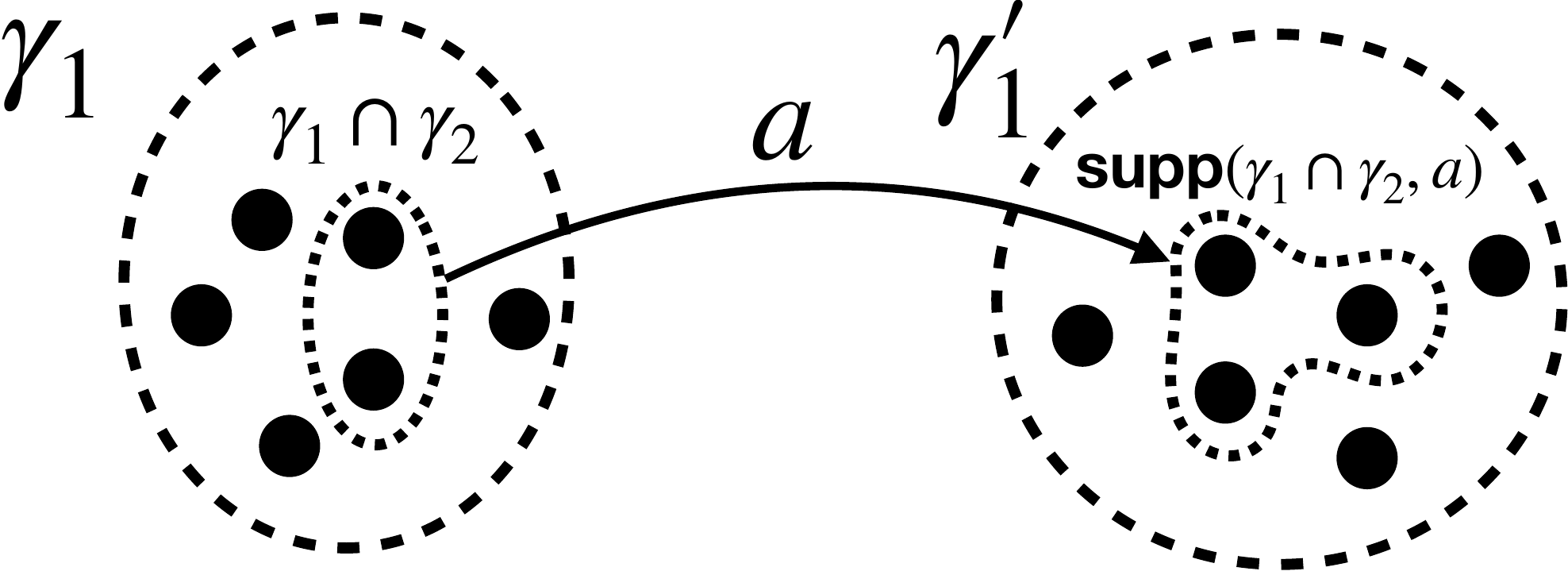}
\caption{A transition between two abstract states $\gamma_1,\gamma'_1 \in \Gamma_1$; black dots represent states in $S$. For every $\s \in \gamma_1 \cap \gamma_2$, we have $\MB(\s) = a$.}
\label{fig:abstraction}
\end{figure}

We describe our construction in which the roles of the players is slightly altered. Consider a plant $\M$ with state space $S$ and actions $A$. The arena of our synthesis game is based on two abstractions $\Gamma_1$ and $\Gamma_2$ of $S$. While we assume $\Gamma_1$ is provided by a user, the partition $\Gamma_2$ is extracted from the magic book. The arena is $\A = \zug{\Gamma_1, A, \Gamma_2, \delta}$, where $\delta$ is defined below. Suppose that the token is placed on $\gamma_1 \in \Gamma_1$ (see Fig.~\ref{fig:abstraction}). Intuitively, the actual location of the plant is a state $\s \in S$ with $\s \in \gamma_1$. \PT moves first and chooses a set $\gamma_2 \in \Gamma_2$ such that $\gamma_1 \cap \gamma_2 \neq \emptyset$. Intuitively, a \PT action reveals that the actual state of the plant is in $\gamma_1 \cap \gamma_2$. \PO reacts by choosing an action $a \in A$. We denote by $\supp(\gamma_1 \cap \gamma_2, a)$ the set of possible next locations the plant can be in, thus $\supp(\gamma_1 \cap \gamma_2, a) = \set{\s': \exists \s \in \gamma_1 \cap \gamma_2 \text{ with } \s' \in \supp(\s, a)}$. Then, the next state in the game according to $\delta$ is the minimal-sized set $\gamma'_1 \in \Gamma_1$ such that $\supp(\gamma_1 \cap \gamma_2, a) \subseteq \gamma'_1$.

Suppose for ease of presentation that the magic book is a decision tree $\T$, and the construction easily generalizes to forests. Recall that a state $\s \in S$ produces a unique path $\eta = \path(\T, \s)$, which corresponds to sequence of predicates $\varphi_1,\ldots, \varphi_n$, and $\pred(\eta) = \bigwedge_{1 \leq i \leq n} \varphi_i$. We define $\Gamma_2 = \set{\pred(\eta) : \eta \text{ is a path in } \T}$. Let $\eta$ be a path in $\T$ and $\gamma_2 \in \Gamma_2$ the corresponding predicates. For ease of notation, we abuse notation and refer to $\gamma_2$ as the set of states in $S$ who produce the path $\eta$ in $\T$. An immediate consequence of the construction is the following. 
\begin{lemma}
\label{lem:distinguish}
For every $\gamma_2 \in \Gamma_2$ there is $a \in A$ such that $\MB(\s)=a$, for all $\s \in \gamma_2$.
\end{lemma}

In the following lemma we formalize the intuition that \PT over-approximates the plant. It is not hard, given a \PO strategy $f$, to obtain a policy $\pi(f)$ that follows it. For $\s \in S$, we use $\gamma_1(\s) \in \Gamma_1$ and $\gamma_2(\s) \in \Gamma_2$ to denote the unique abstract set that $\s$ belongs to.
%short Consider a \PO strategy $f$, we denote by $\pi(f)$ the corresponding policy in $\M$, which is defined as follows. For simplicity, we assume that $f$ is positional, thus it is a function from $f: \Gamma_1 \times \Gamma_2 \rightarrow A$, and it is not hard to extend to non-positional strategies by induction on the play. Since both $\Gamma_1$ and $\Gamma_2$ partition $S$, for every $\s \in S$, there is a unique pair $\gamma_1,\gamma_2$ with $\s \in \gamma_1 \cap \gamma_2$. Given $\s \in S$, let $\gamma_1, \gamma_2$ be the unique pair that $\s$ belongs to. Then, the action that $\pi(f)$ prescribes is $f(\gamma_1, \gamma_2)$.

\begin{lemma}
\label{lem:synth-apprx}
Let $f$ be a \PO strategy. Consider a trace $\sigma = \sigma_1,\sigma_2,\ldots \in L_{\pi(f)}(\M)$. Then, there is a \PT strategy $g$ such that $play(f, g) = \gamma_1(\sigma_1), \gamma_1(\sigma_2), \ldots$.
\end{lemma}
\begin{proof}
We define $g$ inductively so that for every $n \geq 1$, the $n$-th vertex of $play(f, g)$ is $\gamma_1(\sigma_n)$. Suppose the invariant holds for $\sigma_n$. \PT chooses $\gamma_2(\sigma_n)$. The definition of $\delta$ implies that the invariant is maintained, thus $\sigma_{n+1} \in \delta\big(\gamma_1(\sigma_n), f(\gamma_1(\sigma_n), \gamma_2(\sigma_n))\big)$.
\end{proof}

We note that the converse of Lemma~\ref{lem:synth-apprx} is not necessarily correct, thus \PT strictly over-approximates the plant. Indeed, suppose that the token is placed on $\gamma_1$, \PT chooses $\gamma_2$, \PO chooses $a \in A$, and the token proceeds to $\gamma'_1$. Intuitively, the plant state was in $\gamma_1 \cap \gamma_2$ and thus should now be in $\supp(\gamma_1 \cap \gamma_2, a)$. In the subsequent move, however, \PT is allowed to choose any $\gamma'_2$ with $\gamma'_1 \cap \gamma'_2 \neq \emptyset$, even one that does not intersect $\supp(\gamma_1 \cap \gamma_2, a)$.

\subsection{Following expert advice}
In this section, we abstain from solving the problem of finding a correct and optimal controller; a problem that is computationally hard for explicit systems, not to mention symbolically-represented systems like the ones we consider. Instead, in order to add performance consideration to synthesis, we think of the wizard as an authority in terms of performance and solve the (hopefully simpler) problem of constructing a correct controller that follows the wizard's actions as closely as possible. We use the magic book as a proxy for the wizard and assume that following its actions most of the time results in favorable performance.

The game arena is constructed as in the previous section. \PO's goal is to ensure that a given specification $\Omega$ is satisfied while optimizing a quantitative objective that we use to formalize the notion of ``following the magic book''. For simplicity, we consider finite paths, thus $\Omega \subseteq \Gamma_1^*$, and the definitions can be generalized to infinite plays. By Lem.~\ref{lem:distinguish}, every \PT action $\gamma_2 \in \Gamma_2$ corresponds to a unique action in $A$, which we denote by $a(\gamma_2) \in A$. We think of \PT as ``suggesting'' the action $a(\gamma_2)$ since for every $\s \in \gamma_2$, we have $\MB(\s) = a(\gamma_2)$. To motivate \PO to use $a(\gamma_2)$, when he ``accepts'' the suggestion and chooses the same action, he obtains a reward of $1$ and otherwise he obtains no reward. Then, \PO's goal in the game is to maximize the sum of rewards that he obtains.

We formalize the guarantees of the controller $\pi(f)$ that we synthesize w.r.t. an optimal strategy $f$ for \PO. Intuitively, the payoff that $f$ guarantees in the game is a lower on the number of times $\pi(f)$ agrees with the magic book in any trace of the plant.
Let $f$ and $g$ be two strategies for the two players. We use $\texttt{Score}(f,g)$ to denote the payoff of \PO in the game. When $play(f,g) \notin \Omega$, we set $\texttt{Score}(f,g) = \infty$, thus \PO first tries to ensure that $\Omega$ holds. If $play(f,g) \in \Omega$, the score is the sum of rewards in $play(f,g)$. We assign a score to $\pi(f)$ in a path-based manner. Let $\sigma = \sigma_1,\ldots, \sigma_n \in L_{\pi(f)}(\M)$. For every $1 \leq i \leq n$, we issue a reward of $1$ if $\pi(f)(\sigma_i) = \MB(\sigma_i)$, and we denote by $\texttt{Agree}(\pi(f), \sigma)$, the sum of rewards, which represents the sum of states in which $\pi(f)$ agrees with $\MB$ throughout $\sigma$. The following theorem follows from Lem.~\ref{lem:synth-apprx}.

\begin{theorem}
Let $f^*$ be a strategy that achieves $x^* = \max_f \min_g \texttt{Score}(f, g)$. If $x^* < \infty$, then $\pi(f)$ is correct w.r.t. $\Omega$. Moreover, for every $\sigma \in L_{\pi(f^*)}(\M)$ we have $\texttt{Agree}(\pi(f^*), \sigma) \geq x^*$.
\end{theorem}

\subsection{Multi-agent synthesis}
\label{sec:multiagent}
In this section, we design a controller in a multi-agent setting, where traditional synthesis is unrealizable and thus does not return any controller. 

For ease of presentation, we focus on two agents, and the construction can be generalized to more agents in a straightforward manner. We assume that the set of actions $A$ is partitioned between the two agents, thus $A = A_1 \times A_2$. In each step, the players simultaneously select actions, where for $i \in\set{1,2}$, \PLi selects an action in $a_i \in A_i$. As before, the joint action determines a probability distribution on the next state according to $\delta$. Our goal is to find a controller for Agent~$1$ that satisfies a given specification $\Omega$ no matter how \PT plays. 

\begin{example}
Suppose that the grid has two means of transportation: a bus (Agent~$1$) and a taxi (Agent~$2$). We are interested in synthesizing a bus controller for the specification ``travel between two stations while not hitting the taxi''. If one models the taxi as an adversary, the specification is clearly not realizable: the taxi parks in one of the targets so that the bus cannot visit it without crashing into the taxi.
\end{example}

We assume that Agent~$2$ is operating according to a magic book. As in the previous section, we require an abstraction $\Gamma_1$ such that $\Omega \subseteq \Gamma_1^\omega$ and the abstraction $\Gamma_2$ is obtained from the magic book. We construct a game arena as in Section~\ref{sec:game} and \PO wins an infinite play iff it satisfies $\Omega$. 

The way the magic book is employed here is that it restricts the possible actions that \PT can take. Going back to the taxi and bus example, at a state $\s \in S$, \PT essentially chooses how to move the taxi. Suppose the token is placed on $\gamma_1 \in \Gamma_1$. \PT cannot choose to move the taxi in any direction; indeed, he can choose $a_2 \in A_2$ only when there is a state $\s \in \gamma_1$ such that $\MB(\s) = a_2$. The following theorem is an immediate consequence of Lem.~\ref{lem:synth-apprx}.

\begin{theorem}
Let $f$ be a {\em winning} strategy: for every $g$, $play(f, g)$ satisfies $\Omega$. Then, $L_\pi(f)(\M) \subseteq \Omega$.
\end{theorem}

In Remark~\ref{rem:sound} we discuss the guarantees on the magic book that are needed to assume that Agent~$2$ operates according to a wizard rather than a magic book.

\begin{figure}[t]
\centering
\includegraphics[height=4.5cm]{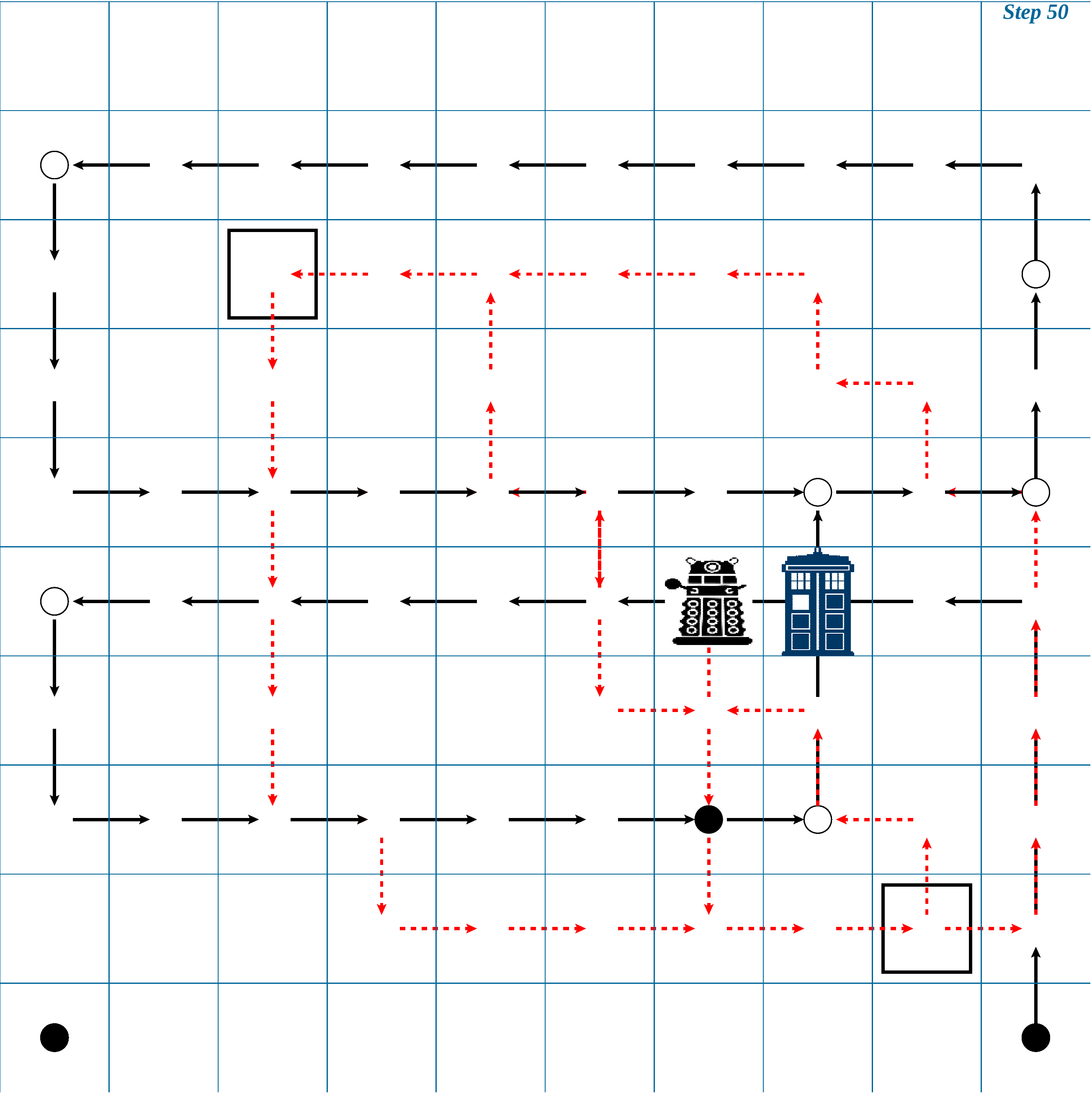}
\caption{Snapshot of step 50 of the simulation. A bus controlled by a synthesized controller (depicted in red dashed arrows and travelling between two square stations) shares the grid with a taxi that is controlled by a magic-book (depicted with black arrows).}
%black squares are the stations between which the bus (the alien) commutes. The trace of the taxi is depicted with black arrows, the trace of the bus is depicted with red dashed arrows.}
\label{fig:bus}
\end{figure}

\section{BMC Based on Magic Books}
In this section, we describe a bounded-model-checking (BMC) \cite{BC+03} procedure that is based on a tree-based magic book. We use our procedure in verification and as an explainability tool to increase the trustworthiness of the wizard before outputting it as the controller for the plant.

\begin{definition}[Bounded model checking] Given a plant $\M$ with state space $S$, a specification $\Omega$, a bound $\ell \in \Nat$, and a policy $\pi$, output a run of length $\ell$ in $L_\pi(\M) \cap \Omega$ if one exists.
\end{definition}

BMC reduces to the satisfiability problem for {\em satisfiability modulo theories} (SMT), where the goal is, given a set of constraints over a set of variables $X$, either find a satisfying assignment to $X$ or return that none exists. We are interested in solving BMC for wizards, i.e., finding a path in $L_\Wiz(\M) \cap \Omega$. However, as can be seen in the proof of Thm.~\ref{thm:BMC} below, the SMT program needs to simulate the execution of the wizard, thus it becomes both large and challenging (due to the \texttt{max} operator) for standard SMT solvers. Instead, we solve BMC for magic books to find a path $\eta \in L_\MB(\M) \cap \Omega$. Since $\MB$ is a good approximation for $\Wiz$, we often have $\eta \in L_\Wiz(\M) \cap \Omega$.

\begin{theorem}
\label{thm:BMC}
BMC reduces to SMT. Specifically, given a plant $\M$ with states $S$, a specification $\Omega \subseteq S^*$, a policy $\pi$ given as a tree-based magic book, and a bound $\ell$, there is an SMT formula whose satisfying assignments correspond to paths of length $\ell$ in $L_\pi(\M) \cap \Omega$.
\end{theorem}
\begin{proof}
The first steps of the reduction are standard. Consider a policy $\pi$ and a bound $\ell \in \Nat$. The variables consist of {\em state variables} $X_0,\ldots, X_\ell$ and {\em action variables} $Y_0, \ldots, Y_{\ell-1}$. We add constraints so that, for a satisfying assignment $\alpha$, for $0 \leq i \leq \ell$, each $\alpha(X_i)$ corresponds to a state in $S$, and for $0 \leq i \leq \ell-1$, each $\alpha(Y_i)$ corresponds to an action in $A$. Moreover, for $0 \leq i \leq \ell-1$, the constraints ensure that $\alpha(X_{i+1}) \in \supp(\alpha(X_i), \alpha(Y_i))$, thus we obtain a path in $L_\pi(\M)$. 

We consider a specification $\Omega$ that can be represented as an SMT constraint over $X_0,\ldots, X_\ell$ and add constraints so that the path we find is in $L_\pi(\M) \cap \Omega$. 

The missing component from this construction ensures that the action $\alpha(Y^i)$ is indeed the action that $\pi$ selects at state $\alpha(X^i)$. For that, we need to simulate the operation of $\pi$ using constraints. Suppose first that $\pi$ is represented using a decision tree $\T$. For a path $\eta$ in $\T$, recall that $\pred(\eta)$ is the predicate $\varphi_1 \wedge \ldots \wedge \varphi_n$ that is satisfied by every state $\s \in S$ such that $\path(\T, \s) = \eta$. Moreover, recall that each $\varphi_j$ is a predicate over $S$. For $0 \leq i < \ell$, we create a copy of $\pred(\eta)$ using the variables $X_i$ so that it is satisfied iff $\alpha(X_i)$ satisfies $\pred(\eta)$. For $a \in A$, let $\text{paths}(\T,a)$ denote the set of paths in $\T$ that end in the action $a$. We add a constraint that states that if $\bigvee_{\eta \in \text{paths}(\T,a)} \pred(\eta)$ is true at time $i$, then $\alpha(Y_i) = a$. Finally, when $\MB$ is a forest, we need to count the number of trees that vote for each action and set $\alpha(Y_i)$ to equal the action with the highest count. 
\end{proof}

\begin{remark}
{\bf (The size of the SMT program).} 
In the construction in Theorem~\ref{thm:BMC}, as is standard in BMC, we use roughly $\ell$ copies of $\M$, where the size of each copy depends on the representation size of $\M$. In addition, we need a constraint that represents $\Omega$, which in our examples, is of size $O(\ell)$. The main bottleneck are the constraints that represent $\pi$. Each path appears exactly once in a constraint, and we use $\ell+1$ copies of $\pi$, thus the total size of these constraints is $O(\ell \cdot |\pi|)$, where $|\pi|$ is the number of paths in the trees in the forest.
\end{remark}

\begin{example}
\label{ex:SMT}
Recall the description of the plant in Example~\ref{ex:plant} in which a taxi travels in a grid. We illustrate how to simulate the plant using an SMT program. A state at time $i$ is a $2\cdot (k+1)$ tuple of variables $\zug{x^i_0, y^i_0, \ldots, x^i_k, y^i_k}$ that take integer values in $\set{0,\ldots, n}$. The position of the taxi at time $i$ is $(x^i_0, y^i_0)$ and the position of Passenger~$j$ is $(x^i_j, y^i_j)$. The transition function is represented using constraints. For example, the constraint $\big(Y_i = \up\big) \rightarrow \big((x^{i+1}_0 = x^i_0) \wedge (y^{i+1}_0 = y^i_0+1)\big)$ means that when the action $\up$ is taken, the taxi moves one step up. The constraint $\neg \big((x^{i+1}_0 = x^i_j) \wedge (y^{i+1}_0 = y^i_j)\big) \rightarrow \big((x^{i+1}_j = x^i_j) \wedge (y^{i+1}_j = y^i_j)\big)$ means that if Passenger~$j$ is not collected by the taxi at time $i+1$, its location should not change. A key point is that when Passenger~$j$ is collected, we do not constrain his new location, thus we replace the randomness in $\M$ with nondeterminism. 
\end{example}

% and the number of constraints is intuitively the representation size of $\pi$. When $\pi$ is given as a tree-based model, the number of constraints is the number of paths in $\pi$, which we assume is significantly smaller than the size of the NN. Moreover and more importantly, when simulating a tree-based classifier, the constraints use mostly Boolean operations that are considerably easier to handle for an SMT solver than the operations used when simulating a NN.

\paragraph{Verification}
In verification, our goal is to find violations of the wizard for a given specification.

\begin{figure}[t]
\centering
\includegraphics[width=0.49\linewidth,
trim={0 15cm 0 0},clip]{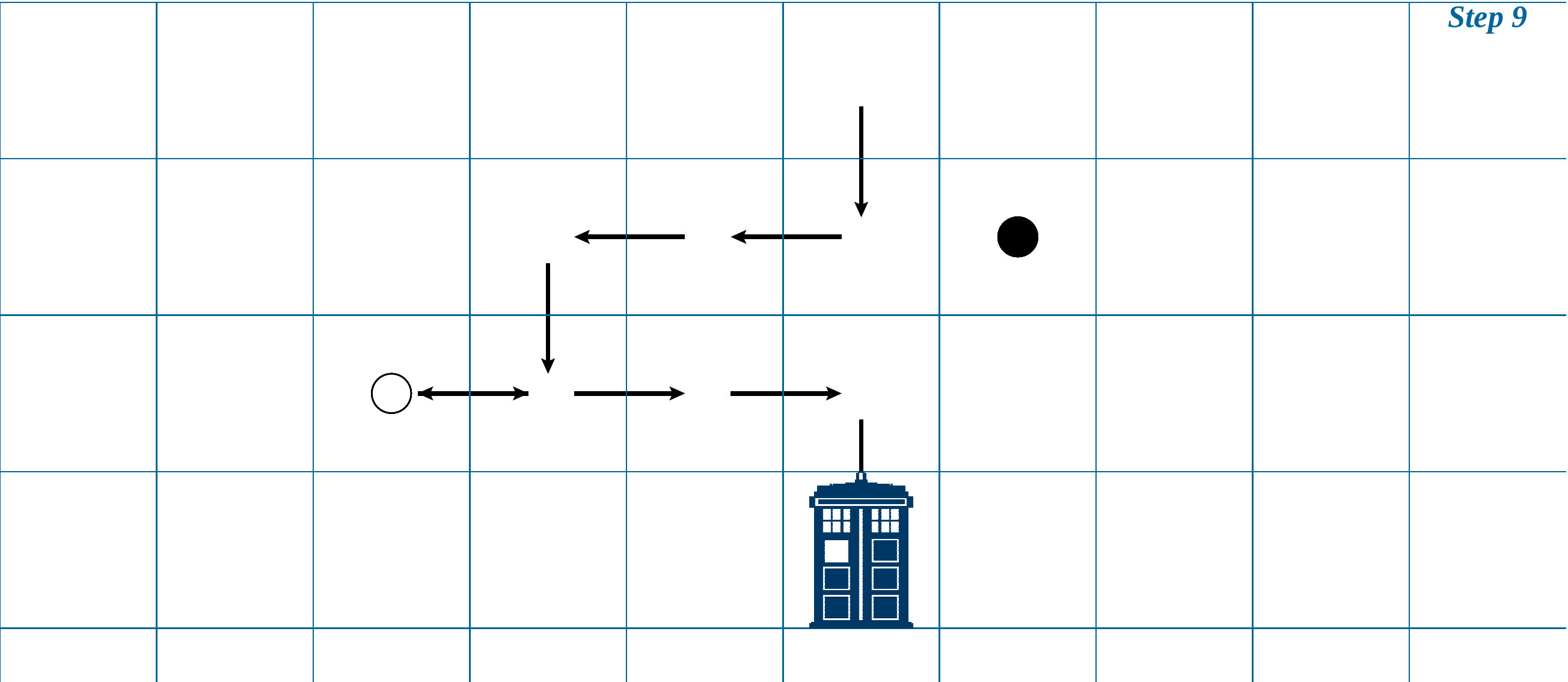}
\hfill
\includegraphics[width=0.49\linewidth
,trim={0 15cm 0 0},clip]{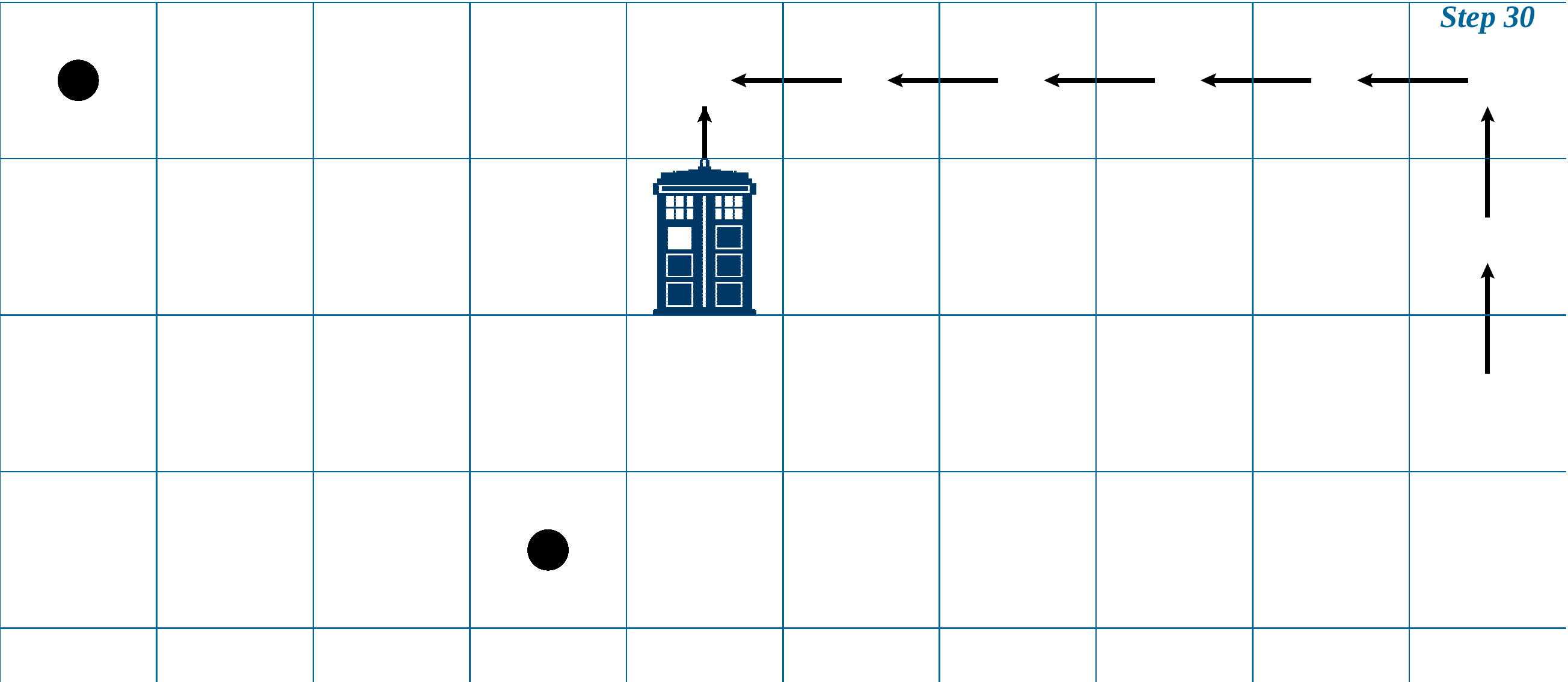}
\caption{Examples found using BMC. Left: a snapshot of step 9 of the simulation showing the closest passenger was not collected first. The passenger collected first is shown as a hollow circle. The passengers not yet collected are shown as filled black circles. Right: a snapshot of step 30 of the simulation of a ``lasso''-shaped trace of the taxi that entered a loop without collecting any passengers.}% The trace is depicted with black arrows.}
\label{fig:XAI-traces}
\end{figure}

\begin{example} 
We show how to express the specification ``the taxi never enters a loop in which no passenger is collected'' as an SMT constraint based on the construction in Example~\ref{ex:SMT}. We simplify slightly and use the constraint $\big(x^\ell_0 = x^0_0 \wedge y^\ell_0 = y^0_0\big)$ that means that the taxi returns to its initial position to close a cycle at the end of the trace. We add a second constraint $\bigwedge_{1 \leq j \leq k} \bigwedge_{1 \leq i \leq \ell} \big(x^0_j = x^i_j \wedge y^0_j = y^i_j\big)$ that means that all passengers stay in their original position throughout the trace. In Fig.~\ref{fig:XAI-traces} (right), we depict a lasso-shaped trace that witnesses a violation of this property.
\end{example}

\begin{remark}[Soundness]
\label{rem:sound}
The benefit of using magic books is scalability, and the draw-back is soundness. For example, when the SMT formula is unsatisfiable for a bound $\ell \in \Nat$, this only means that there are no violations of the {\em magic book} of length $\ell$, and there can still be a violation of the wizard. To regain soundness we would need guarantees on the relation between the magic book and the wizard. An example of a guarantee is that the two functions coincide, thus for every state $\s \in S$, we have $\Wiz(\s) = \MB(\s)$. However, if at all possible, we expect such a strong guarantee to come at the expense of a huge magic book, thus bringing us back to square one. We are more optimistic that one can find small magic books with approximation guarantees. For example, one can define a magic book as a function $\MB: S \rightarrow 2^A$ that ``suggests'' a set of actions rather than only one, and require that for every state $\s \in S$, we have $\Wiz(\s) \in \MB(\s)$. %That is, the magic book suwhen wizard chooses $a \in A$, then $a$ is one of the magic book's suggestions. 
Such guarantees suffice to regain soundness both in BMC and for the synthesis application in Section~\ref{sec:multiagent}. We leave for future work obtaining such magic books.
\end{remark}

%\begin{remark}
%Some weaknesses of BMC are carried over to our approach. First, in the case that there is no $\sigma$ of length $\ell$ in $L_\MB(\M) \cap \Omega$, we increase $\ell$ and repeat. As in BMC, there is a challenge in deciding when to terminate the algorithm, and the methodology developed in BMC can be used here as well (see \cite{handbookMC}). Second, the technique can only verify specifications that have short counterexamples. Fortunately, many specifications, including safety and liveness, have this property. 
%\end{remark}

\paragraph{Explainability}
We illustrate how BMC can be used as an XAI tool. BMC allows us to find corner-case traces that are hard to find in a manual simulation and the individual traces can serve as explanations. For example, in Fig.~\ref{fig:XAI-traces} (left), we depict a trace that is obtained using BMC for the property ``the first passenger to be collected is not the closest''.

%Intuitively, BMC allows us to find many interesting examples that are hard to collect from arbitrary simulations. For example, we can use the framework to find traces in which ``the first passenger to be collected is not the closest to the taxi''. These traces can serve as explanations by themselves (see Fig.~\ref{fig:XAI-traces} on the left). 

A second application of BMC is based on gathering a large number of traces. We construct a small human-readable model that explains the decision procedure of the wizard. We note that while the magic book is already a small model that approximates the wizard, its size is way too large for a human to reason about. For us, a small model is one decision tree of depth at most~$4$. Moreover, the magic book is a ``local'' function, its type is from states to actions, whereas a human is typically interested in ``global'' behavior, e.g., which action to take next as opposed to which passenger is collected next, respectively. 

We rely on the user to supply specifications $\Omega_1,\ldots, \Omega_m$. We gather a dataset that consists of pairs of the form $\zug{\s, i}$, for each $1 \leq i \leq m$, where $\s$ is such that when the plant starts at configuration $\s$ under the control of the wizard, then $\Omega_i$ is satisfied.
To find many traces that satisfy $\Omega_i$, we iteratively call an SMT solver. Suppose it finds a trace $\eta \in \Omega_i$. Then, before the next call, we add the constraint $\neg \eta$ to the SMT program so that $\eta$ is not found again. In practice, the amortized running time of this simple algorithm is low. One reason is that generating the SMT program takes considerable time, even when comparing to the time it takes to solve it. This running time is essentially amortized over all executions since the running time of adding a single constraint is negligible. In addition, the SMT solver learns the structure of the SMT program and uses it to speed up subsequent executions.

\begin{example}
Suppose we are interested in understanding if and how the wizard prioritizes collecting passengers. We consider the specifications ``Passenger~$j$ is collected first'', for $1 \leq j \leq k$. It can be formalized using the following constraints. The constraint $\bigwedge_{1 \leq i \leq \ell} (x^i_j = x^0_j \wedge y^i_j = y^0_j)$ means that Passenger~$j$ is not collected since it stays in place throughout the whole trace, and we add such a constraint for all but one passenger. The constraint $\neg (x^\ell_j = x^0_j \wedge y^\ell_j = y^0_j)$ means that Passenger~$j$ must have been collected at least once since its final position differs from his initial position. In Fig.~\ref{fig:XAI-tree} we depict a tree that we extract using these specifications.
\end{example}

\section{Experiments}

\paragraph{Setup}
We illustrate our approach using an implementation of the case study that is our running example: a taxi traveling on a grid and collecting passengers. 
We set the size of the grid to be $n=10$ and the number of passengers to $m=3$, thus the state space is almost $10^8$. All simulations were programmed in Python and run on a personal computer with an Intel Core i3-4130 3.40GHz CPU, 7.7 GiB memory runnning Ubuntu. 

\paragraph{Training a wizard using deep RL}
The plant state in our training is a $6$-tuple that, for each passenger, contains the distances to the taxi on both axes. When the taxi collects a passenger, the agent receives a reward of $100$. Multi-objective RL is notoriously difficult because the agent gets confused by the various targets. We thus found it useful to add a ``hint'' when the taxi does not collect a passenger: at time $t >1$, if a passenger is not collected, the agent receives a reward of $\max_{i=1,2,3}\left(1/d_{t+1,i}-1/d_{t,i}\right)$, where $d_{j,i}$, for $j \geq 1$ and $i \in \set{1,2,3}$, is the manhattan distance between the taxi and passenger~$i$ at time $j$. We use the Python library Keras~\cite{chollet2015keras} and the ``Adam'' optimizer~\cite{kingma2014adam} to minimize mean squared error loss. We train a NN with two hidden layers that use a ReLU activation function and with $200$ and $100$ neurons, respectively, and a linear output layer. Each episode consists of $1000$ steps and we train for $2000$ episodes.

\paragraph{Extracting the magic book}
We extract configuration-action pairs from $1000$ episodes of the trained agent. % with the plant each of $1000$ time steps. 
We use Python's scikit-learn library~\cite{scikit-learn} to fit one of the tree-based classification model to the obtained dataset. Table~\ref{tbl:wiz_vs_mb} depicts a comparison between the models and the wizard on $200$ episodes. % each with $1000$ time steps. 
%{\em Approximation accuracy} is the ratio of the time that the two functions agree on their outputs. 
{\em Performance} refers to the total number of passengers collected in a simulation. It is encouraging that small forests with shallow trees (of depth not more than $10$) approximate the wizard well. 
 
\paragraph{Synthesis: Following expert advice}
The specification we consider is ``reach a gas station every $t$ time steps'', for some $t \in \Nat$. Our controllers exhibit performance that is not too far from the wizard: see Table~\ref{tbl:wiz_vs_mb} for the performance with $t=30$ and synthesis based on different tree models (take into account that the wizard does not visit the gas station). We view this experiment as a success: we achieve our goal of synthesizing a correct controller that achieves favorable performance. We point out that since traditional synthesis does not address performance, a controller that it produces visits the gas station every $t$ steps but does not collect any passenger.

\begin{figure}[t]
\centering
\includegraphics[width=.49\linewidth,%height=3.4cm,%width=\linewidth,
trim={0 5cm 0 0},clip]{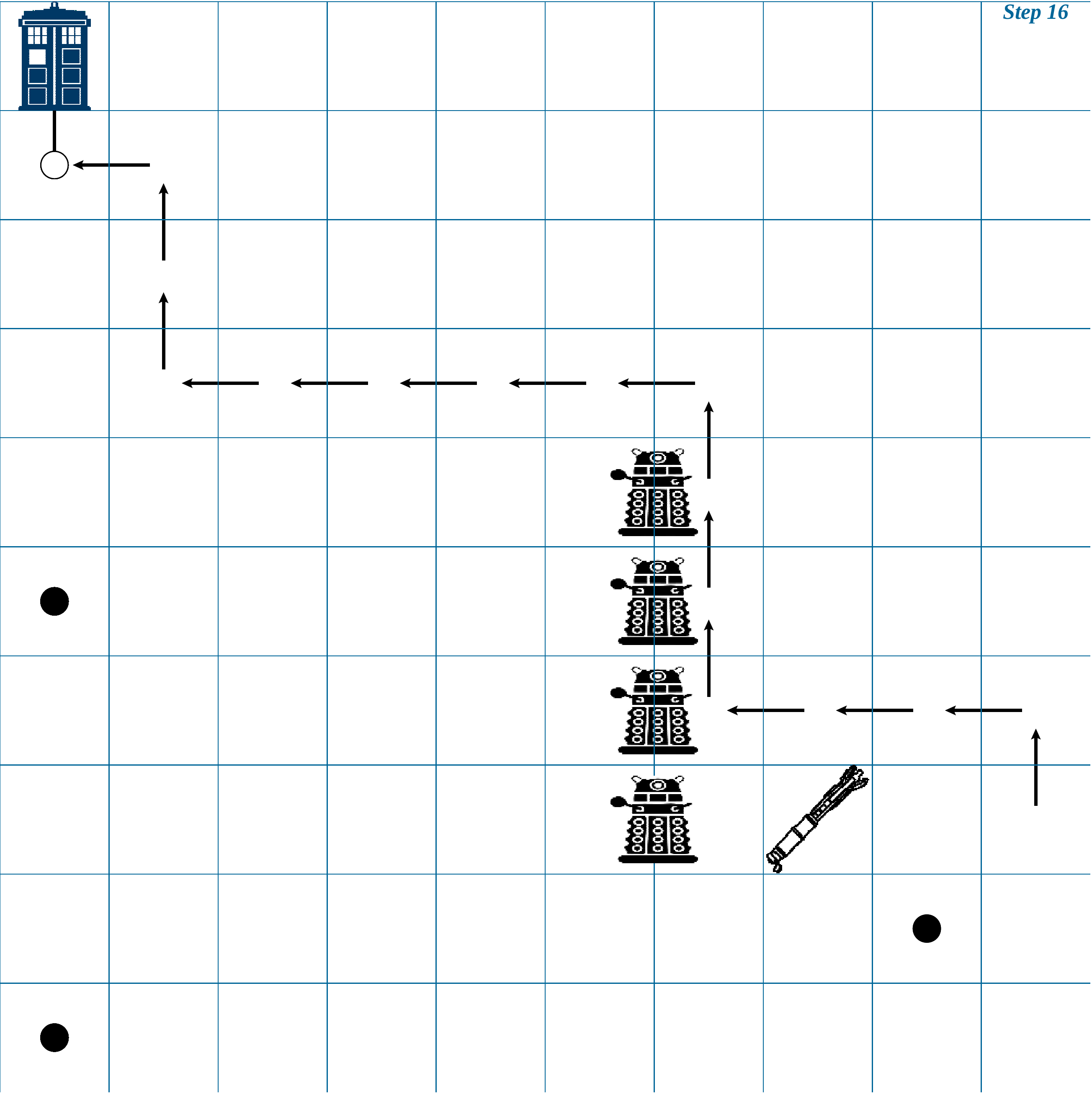}
\hfill
\includegraphics[width=.49\linewidth,%height=3.4cm,%width=.45\textwidth,
trim={0 5cm 0 0},clip]{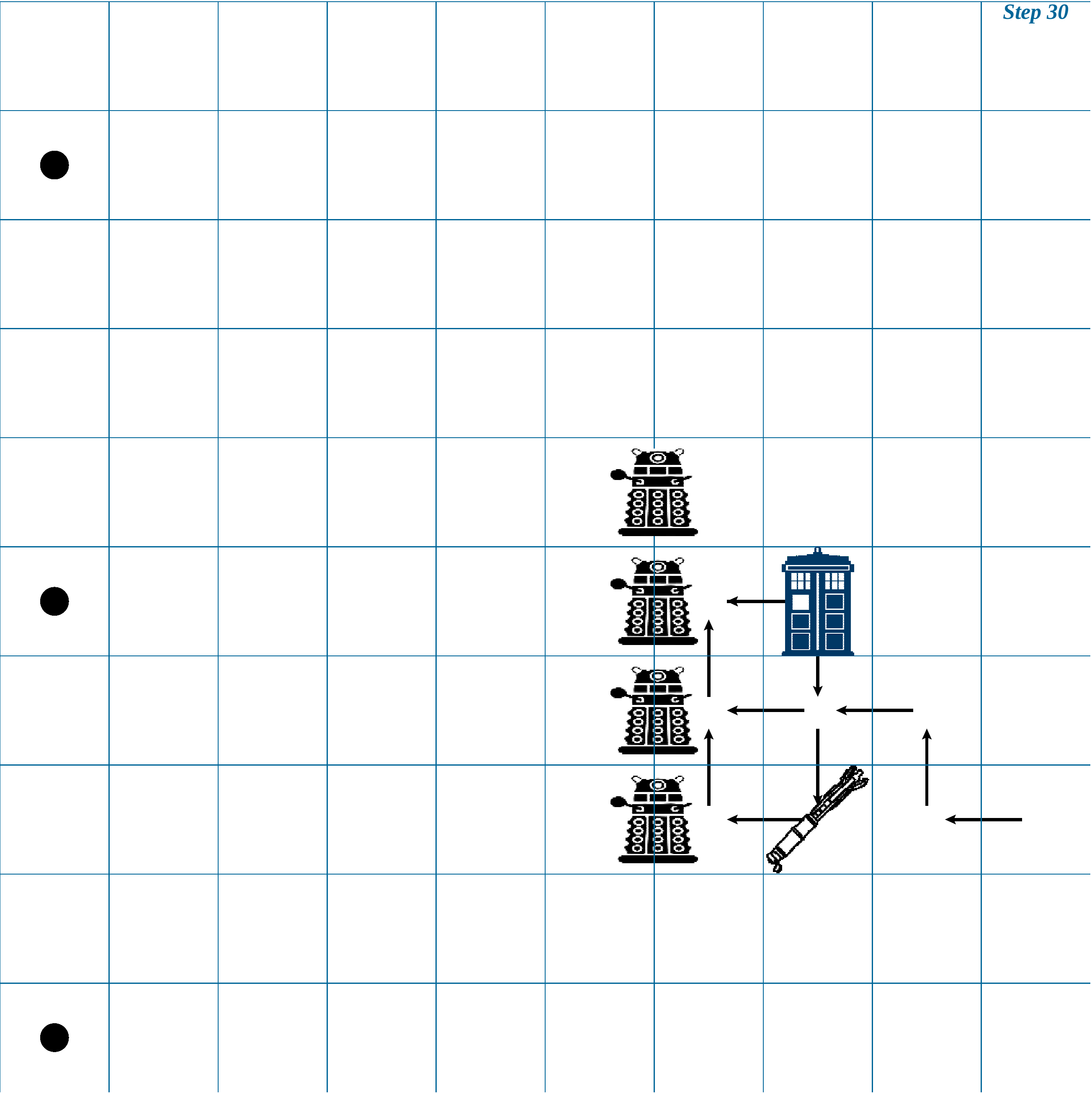}
\caption{Snapshots of simulations showing that a controller, synthesized using a magic-book, crosses a wall (left) whereas a shield-based controller is stuck (right).}
% when facing a wall (aliens) and visiting the gas station (screwdriver): step 16 with synthesis (left); step 30 with shield (right). }The trace is depicted with black arrows.}
\label{fig:synthesis-wall}
\end{figure}

\paragraph{Comparing with a shield-based approach}
A shield-based controller \cite{KAB+17,AB+19} consists of a shield that uses a wizard as a black box: given a plant state $\s$, the wizard is run to obtain $a = \Wiz(\s)$, then $a$ is fed to the shield to obtain $a' \in A$, which is issued to the plant. We demonstrate how our synthesis procedure manages to open up the black-box wizard. In Fig.~\ref{fig:synthesis-wall}, we depict the result of an experiment in which we add a wall to the grid that was not present in training. Crossing a wall is inherently impossible for the shield-based controller since when the wizard suggests an action that is not allowed, the best the shield can do is choose an arbitrary substitute. Our controller, on the other hand, intuitively directs the taxi to areas in the grid where the magic book is ``certain'' of its actions (a notion which is convenient to define when the magic book is a forest). Since these positions are often located near passengers, the taxi manages to cross the wall. 

%short The second experiment is depicted in Fig.~\ref{syn-shield-stuck}. The shield-based controller always attempts to collect the same passenger, but must return to the gas station due to the timeout. Thus, the taxi ends up in a loop with no passengers collected. Our controller directs the taxi to a passenger that would not have been collected first by the magic book.

\begin{table}[t]
    \centering
    \scriptsize{
    \def\d{@{\hspace*{1.7mm}}}
    \begin{tabular}{@{} l | c | c | c | c  @{}}
         Num. of collected passengers & DT(10) & RF(5,6) & xGB(100,10) & Wizard\\
        \hline
        %Approximation accuracy, $\%$ & & $82$ & & $100$  \\ 
        Avg. performance & $147$ & $154$ & $158$ & $159$ \\
        Max. performance & $194$ & $194$ & $190$ & $200$\\
%        Training accuracy, $\%$& & $90$ & &  
        \hline
        Synthesis avg. performance & $122$ & $96$ & -- & --
    \end{tabular}
    }
    \caption{Performances of the wizard compared to three classifiers: decision tree DT(depth), random forest RF(trees, depth), and extreme gradient boosting xGB(trees, depth). Each simulation was ran 10 times for an arbitrary $\bar{s}_0$ and time bound $T=1000$.}
    \label{tbl:wiz_vs_mb}
\end{table}

%average score =  165.9
%maximum score =  178
%average score =  163.3
%maximum score =  189
%average score =  167.9
%maximum score =  185
%average score =  148.9
%maximum score =  184
%average score =  163.2
%maximum score =  178

   % importance_splitting('./rf56', False)
    %importance_splitting('./dt10', True)
    %importance_splitting('./gb10', True)
    %importance_splitting('', False)
    %importance_splitting('./rf510', False)

\paragraph{BMC: Scalability and success rate}
We use the standard state-of-the-art SMT solver Z3~\cite{MouraB08} to solve BMC. In Table~\ref{tab:scalability}, we consider the following specifications for XAI: ``Passenger~$i$ is collected first and at time $\ell$, even though it is not closest'', where $\ell$ is the bound for BMC and for $i \in \set{1,2,3}$. We perform the following experiment $10$ times and average the results. We run BMC to collect $250$ traces. We depict the amortized running time of finding a trace, i.e., the total running time divided by $250$. Recall that the traces witness the magic book. We count the number of traces out of the $250$ that also witness the wizard, and depict their ratio. We find both results encouraging: finding a dataset of non-trivial witness traces of the wizard is feasible.

\begin{table}[t]
    \centering
    \scriptsize{
    \def\d{@{\hspace*{1.7mm}}}
    \begin{tabular}{@{} l | c | c | c | c | c | c | c  @{}}
         \multirow{2}{*}{Bound} & \multicolumn{2}{c|}{Passenger 1} & \multicolumn{2}{c|}{Passenger 2} & \multicolumn{2}{c|}{Passenger 3} \\
%        \hline
        & runtime & succ. ratio & runtime & succ. ratio & runtime & succ. ratio\\
        \hline
        6 & 0.26 s & 82.8 \% &0.25 s & 85 \% & 0.25 s & 81.2 \% \\
        \hline
        7 & 0.30 s & 76.9 \% & 0.30 s & 87.2 \% & 0.30 s & 84.2 \%\\
        \hline
        8 & 0.37 s &85.2 \% & 0.36 s & 89.9 \% & 0.37 s & 88.7 \% \\
        \hline
        9 & 0.44 s & 85.1 \% & 0.47 s & 82.2 \% & 0.49 s & 79.7 \% \\
        \hline
        
        %Approximation accuracy, $\%$ & & $82$ & & $100$  \\ 
%        Avg. performance & $147$ & $154$ & $158$ & $159$ \\
  %      Max. performance & $194$ & $194$ & $190$ & $200$\\
%        Training accuracy, $\%$& & $90$ & &  
  %      \hline
   %     Synthesis avg. performance & $122$ & $96$ & -- & --
    \end{tabular}
    }
    \caption{Results for BMC with bounds $6-9$ using a forest with $5$ trees of depth $10$ as a magic book. The amortized running times for obtaining a trace, over 250 traces, and the ratio of traces that are witnesses for the wizard.}
    \label{tab:scalability}
\end{table}

\paragraph{Wizard-based BMC} 
We implemented a BMC procedure that simulates the wizard instead of the magic book and ran it using Z3. We observe extremely poor scalability: an extremely modest SMT query to find a path of length $2$ timed-out at $20$min, and even when the initial state is set, the running time is $4.51$min! 

\begin{figure}[t]
\centering
\includegraphics[width=\linewidth]{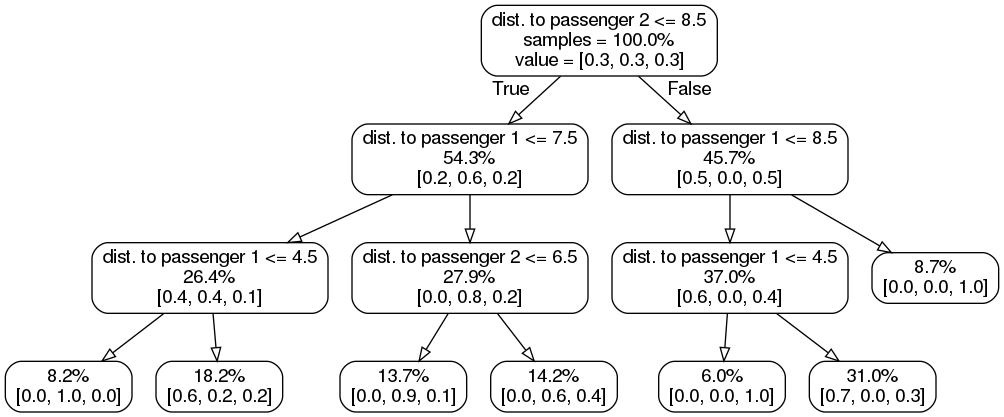}
\caption{A decision tree extracted from a $1200$-sample dataset, obtained using BMC, of the form $\zug{\s, i} \in S\times\set{1,2,3}$, where passenger~$i$ is collected first from the initial state $\s$.}
\label{fig:XAI-tree}
\end{figure}

\paragraph{BMC: Verification and Explainability}
For verification, we consider the specifications ``the taxi never hits the wall'' and ``the taxi never enters a loop in which no passenger is collected''. Even though violations of these specifications were not observed in numerous simulations, we find counterexamples for both (see a depiction for the second property in Fig.~\ref{fig:XAI-traces} on the right). We illustrate explainability with the property ``the closest passenger is not collected first'' by depicting an example trace for it in Fig.~\ref{fig:XAI-traces} on the left. In Fig.~\ref{fig:XAI-tree}, we depict a decision tree, obtained from a dataset consisting of $1200$ examples, as an attempt to explain when the wizard chooses to collect passenger~$i$ first, for $i \in \set{1,2,3}$.

\section{Discussion}
\label{sec:disc}
In this work, we address the controller-design problem using a combination of techniques from formal methods and machine learning. The challenge in this combination is that formal methods struggle with the use of neural networks (NNs). We bypass this difficulty using a novel procedure that, instead of reasoning on the NN that deep RL trains (the wizard), extracts from the wizard a small model that approximates its operation (the magic book). We illustrate the advantage of using the magic book by tackling problems that are currently out of reach for either formal methods or machine learning separately. Specifically, to the best of our knowledge, we are the first to incorporate a magic book in a reactive synthesis procedure thereby synthesizing a stand-alone controller with performance considerations. Second, we use a magic-book based BMC procedure as an XAI tool to increase the trustworthiness of the wizard.

%The first application is based on a BMC procedure that finds executions of the plant that are of interest to the designer and are hard to find manually. In the second application we synthesize a correct-by-design stand-alone controller that optimizes performance by following wizard advice. 

We list several directions for future work. %The decision-trees we extract have no guarantees.
%While we use tree-based models as magic books, the concept of a magic book is general. 
We find it an interesting and important problem to extract magic books with provable guarantees (see Remark~\ref{rem:sound}). 
%In terms of extracting magic books, we find it interesting to show extractions of magic books in other domains as well as extracting a magic book with provable guarantees (see Remark~\ref{rem:sound}). 
%More generally, this work uses XAI techniques, which are typically aimed at humans, to explain the operation of a NN to a machine (formal methods). We find this to be a general and promising direction to develop trustworthy systems. 
Another line of future work is finding other domains in which magic books can be extracted and other applications for magic books. One concrete domain is in speeding up solvers (e.g., SAT, SMT, QBF, etc). Recently, there are attempts at replacing traditional engineered heuristics with learned heuristics~(e.g, \cite{SKM19,LRSL20}). This approach was shown to be fruitful in \cite{YP19}, where an RL-based SAT solver performed less operations than a standard SAT solver. However, at runtime, the SAT solver has the upper hand since the bottleneck becomes the calls to the NN. We find it interesting to use a magic book instead of a NN in this domain so that a solver would benefit from using a learned heuristic without paying the cost of a high runtime.

Our synthesis procedure is based on an abstraction of the plant. In the future, we plan to investigate an iterative refinement scheme for the abstraction. Refinement in our setting is not standard since it includes a quantitative game (e.g.,~\cite{AK12}), and more interesting, there is inaccuracy introduced by the magic book and wizard. Refinement can be applied both to the process of extracting the decision tree from the NN as well as improving the performance of the wizard using training.

\section{Acknowledgements}
This research was supported in part by the Austrian Science Fund (FWF) under grant Z211-N23 (Wittgenstein Award).

\bibliographystyle{IEEEtran}
\bibliography{wizard}
\end{document}